# Dark Energy Detected with Supervoids and Superclusters

Benjamin R. Granett, Mark C. Neyrinck, & István Szapudi

*Institute for Astronomy, University of Hawaii, 2680 Woodlawn Dr, Honolulu, HI 96822, USA*

**The observed apparent acceleration of the universe[1,2,3] is usually attributed to negative pressure from a mysterious dark energy. This acceleration causes the gravitational potential to decay, heating or cooling photons travelling through crests or troughs of large-scale matter density fluctuations. This phenomenon, the late-time integrated Sachs-Wolfe (ISW) effect[4,5], has been detected, albeit at low significance, by cross-correlating various galaxy surveys with the Cosmic Microwave Background (CMB)[6,7,8]. Recently, the best evidence has come from the statistical combination of results from multiple correlated galaxy data sets[9,10]. Here we show that vast structures identified in a galaxy survey project an image onto the CMB; stacking regions aligned with superclusters produces a hot spot, and supervoids, a cold spot. At over $4\sigma$, this is the clearest evidence of the ISW effect to date. For the first time, our findings pin the effect to discrete structures. The ISW signal from supervoids and superclusters can be combined with other cosmological probes to constrain dark energy and cosmological parameters. In addition, our findings make it more plausible that the extreme Cold Spot[11] and other anomalies in the CMB are caused by supervoids[12].**

The observed dimming of distant supernovae suggests that vacuum or dark energy significantly contributes to the total cosmological energy density. The best-fitting value of this contribution is $10^{120}$ times smaller than the one suggested by quantum field theory[13], possibly making dark energy the most profound puzzle of contemporary physics. Cosmology research is focused on detecting and characterizing dark energy[13] through its effect on the expansion of the Universe. Geometric probes use standard candles (such as supernovae), and standard rods (such as the imprint of a physical scale, the sound horizon, on the distribution of galaxies or the CMB). Dynamic probes test the expansion history through its effect on the growth of small fluctuations. The integrated Sachs-Wolfe (ISW) effect, also known as the Rees-Sciama[5] effect, is unique in that it directly probes dark energy[14].

The cross-correlation of galaxies with the CMB temperature is a probe of the ISW effect, but the expected signal-to-noise of the measurement is low primarily because both temperature and galaxy autocorrelations, quite useful for other cosmological estimates, contribute substantially to the noise. Multiple galaxy data sets may be combined, but modelling their covariances introduces uncertainties in the statistical significance of the results[9,10]. To improve the significance of the detection of the ISW effect and to test whether a large fraction of the signal can be attributed to individual sites, we identified the largest over- and under-densities in a galaxy catalogue. These structures are still undergoing gravitational collapse and expansion, respectively. The superclusters we detect are unlikely to be gravitationally bound, which distinguishes them from their constituent clusters. We use the term 'supervoids' for the corresponding large, low-density regions. The physical scale of these gently underdense structures is of order 100 $h^{-1}$ Mpc, above which the universe truly starts to approach homogeneity. The term void usually means a smaller, highly underdense and nonlinear structure lying entirely below the mean density[15], while supervoids might even contain clusters.

We used a sample of 1.1 million Luminous Red Galaxies (LRGs) from the Sloan Digital Sky Survey[16] (SDSS), covering 7500 square degrees about the North Galactic pole. They span a redshift range of 0.4<$z$<0.75, with a median of ~0.5, and inhabit a volume of about 5 $h^{-3}$ Gpc$^3$. LRGs are elliptical galaxies in massive galaxy clusters representing large dark-matter halos[17], and are thought to be physically similar objects across their redshift range[18,19]. This makes them excellent, albeit sparse, tracers of the cosmic matter distribution on scales >~10 Mpc. Our sample was selected from photometric data based on the criteria used in the Mega-Z LRG catalogue[20] (see supplement for details).

We used the CMB temperature map constructed as an inverse-variance weighted combination of the *Wilkinson Microwave Anisotropy Probe* (WMAP) 5-year Q, V and W frequency maps[21], with the foreground galactic-emission maps subtracted from each. Regions within the extended temperature analysis mask (KQ75), which is a conservative Galactic and point source mask, are left out of the analysis. The maps are binned[22] with a pixel size of 7 arcminutes, which oversamples the 30-arcminute full-width, half-max beam. In agreement with previous results[10], we measured a cross-correlation between our two data sets on 1° scales of about 0.7$\mu$K at 2.2$\sigma$.

To find supervoids in the galaxy sample, we used the parameter-free, publicly available ZOBOV[23] (ZOnes Bordering On Voidness) algorithm. For each galaxy, ZOBOV estimates the density and set of neighbours using the parameter-free Voronoi tessellation[24]. Then, around each density minimum, ZOBOV finds density depressions, i.e. voids. We call the ZOBOV-detected objects 'voids,' reserving 'supervoids' for the largest ones that give a significant ISW signal; similarly for clusters. We used VOBOZ[25] to detect clusters, the same algorithm applied to the inverse of the density. (See supplement for details.)

We found 513 voids and 3342 clusters above a 2$\sigma$ significance level, evaluated by comparing their density contrasts to those of voids and clusters in a uniform Poisson point sample. There are so many structures because of the high sensitivity of the Voronoi tessellation; almost all of them are spurious, arising from discreteness noise. We thus used only the 50 voids and 50 clusters with the highest significance in our analysis. Structures at these thresholds are unlikely to arise from discreteness noise at about a 4$\sigma$ level. We discarded any structures with over 10% overlap with cut out holes (due to bright stars, etc.) in the survey, and within 2.5° of the footprint boundary. Figure 1 shows the locations of these supervoids and superclusters.

Figure 2 shows a stack image built by averaging the regions on the CMB surrounding each object. The CMB stack corresponding to supervoids shows a cold spot of -11.3µK with 3.7$\sigma$ significance, while that corresponding to superclusters shows a hot spot of 7.9µK with 2.6$\sigma$ significance, assessed in the same way as for the combined signal, described below. The characteristic size of the detections is about 4°, consistent with the theoretical expectation[9] that the ISW effect peaks around the spherical-harmonic multipole of $l$=20. At the median redshift of the objects, this corresponds to a scale of 100 $h^{-1}$ Mpc.

The most plausible interpretation of the above picture is that the structures we found correspond to supervoids and superclusters affecting the CMB through the ISW effect. To assess the significance of our detection, we subtracted the supervoid image from the supercluster image. We averaged the temperature within 4° of the centre, and then subtracted the mean temperature in a ring of the same area around it. This is a simple top-hat compensated filter, which is insensitive to CMB fluctuations on scales larger than the object detected; for an uncompensated filter, these fluctuations would constitute a significant source of noise. What is the likelihood that our results are due to random fluctuations? To estimate that, we performed two sets of 1000 Monte Carlo simulations. First, we generated random positions of voids and clusters within the survey and stacked the corresponding areas of the actual CMB map. This models the errors given the observed CMB sky and foreground subtraction, but might not properly account for any covariance due to the actual configuration of voids and clusters. Second, we generated model CMB skies smoothed to WMAP resolution and repeated our analysis on these with the actual void and cluster configurations observed in the catalogues. We find that these two approaches produce identical distributions consistent with Gaussians, and with standard deviations within 2% of each other. The hypothesis that the signal arose from random fluctuations is excluded at the 4.4$\sigma$ level, a 1:200000 chance. Our final mean signal with errors is 9.6±2.2$\mu$K.

We checked the robustness of our findings by varying the number of objects (30-70 each), and the filter sizes. The signal is diluted to 2.8$\sigma$ in a stack of the top 70 voids and 70 clusters; however, with 30 the signal remains above 4$\sigma$. Changing the filter size between 3-5° results in various detection significances between 3.5-4.5$\sigma$. The summary of these tests is presented in Tables 2 and 3 of the supplement. We conclude that our detection of the ISW effect and our estimation of its significance are robust with respect to details of our procedure.

The consensus in the literature is that detecting the ISW effect signals the presence of dark energy in our flat, dark energy dominated ($\Lambda$CDM) Universe. The non-linear Rees-Sciama effect in a flat universe without dark energy is expected to be about an order of magnitude smaller than the linear ISW in standard $\Lambda$CDM[26,27]. To estimate the expected effect from ISW, we measured the signal that the Millennium cosmological *N*-body simulation produces. In this particular volume, which is large enough for 1 or 2 supervoids and superclusters, we confirmed that the linear part of the ISW signal dominates over higher-order effects. However, the Millennium volume gives a signal that is ~2$\sigma$ lower than what we observed in our CMB stack (see supplement). Though we only expect these numbers to agree to within an order of magnitude, we note that most previous ISW measurements are also somewhat higher than the predicted signal in a ΛCDM cosmology[9]. While more theoretical studies are needed to turn our detection into precision constraints on cosmological parameters, we interpret our image as the ISW effect on the CMB caused by the decaying of potentials in an accelerating universe with dark energy.

Based on our detection, we speculate that low-redshift supervoids and superclusters might explain some or even all of the anomalies observed on the CMB[12,28]. At low to moderate significance, these features include a 5° 70μK Cold Spot, the North-South power asymmetry, the low quadrupole moment, and the alignment of low multipoles. Additionally, $f_{\rm nl}$, a measure of non-Gaussianity on the CMB, has been estimated to be positive at low significance in WMAP. This indicates a CMB temperature distribution that is slightly skewed towards low temperatures, as predicted by a small nonlinear ISW effect that enhances supervoid signals over superclusters[29]. We indeed find somewhat stronger cold spots, and although the difference is not statistically significant, its consistency with the above picture is intriguing.

For supplementary information, please see pages 9-17, as well as http://ifa.hawaii.edu/cosmowave/supervoids/.

**Acknowledgements** We thank Adrian Pope for useful discussions and help with the SDSS masks and catalogue. The authors were funded by NSF and NASA.

Authors can be reached for correspondence and requests by e-mail (granett@ifa.hawaii.edu, neyrinck@ifa.hawaii.edu, szapudi@ifa.hawaii.edu ).

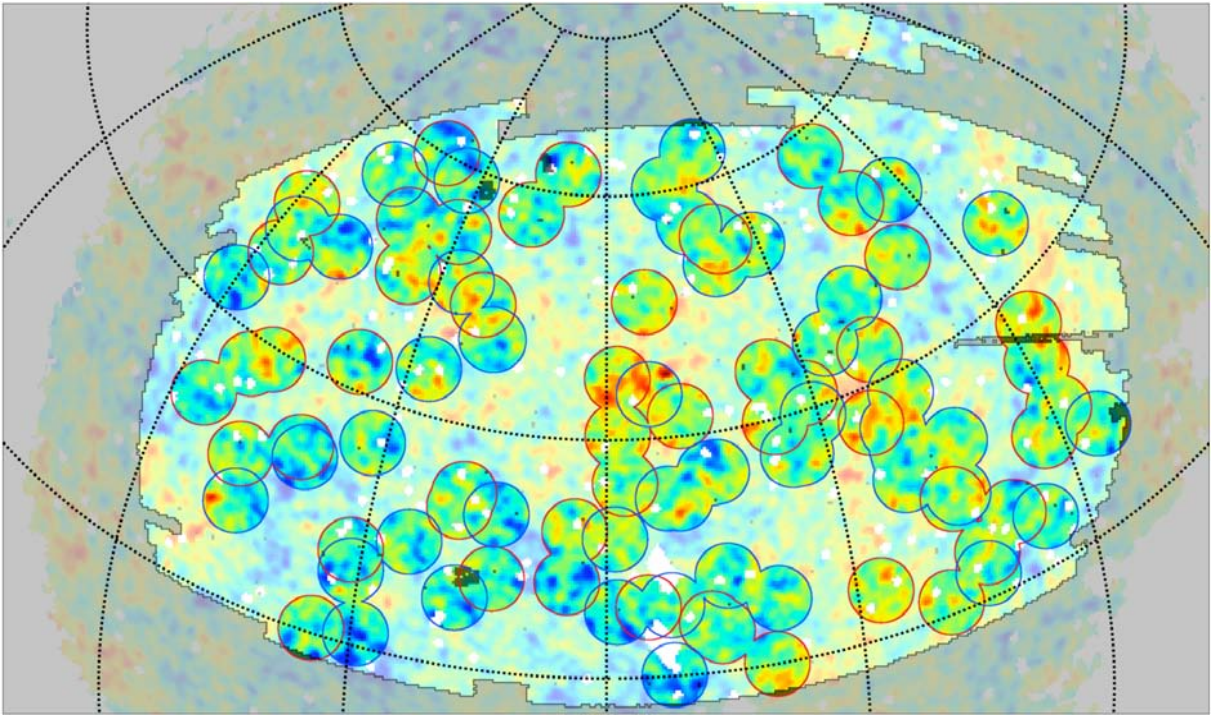

Figure 1: A map of the microwave sky over the SDSS area. The supervoids and superclusters used in our analysis are highlighted and outlined at a radius of 4°, blue for supervoids and red for superclusters. The compensated filter we use in our analysis approximately corrects for the large-angular-scale temperature variations that are visible across the map. The SDSS DR6 coverage footprint is outlined. Holes in the survey, *e.g.* due to bright stars, are displayed in black. Additionally, the WMAP Galactic foreground and point source mask is plotted (white holes). The disk of the Milky Way, which extends around the left and right border of the figure, is also masked. The map is in a Lambert azimuthal equal-area projection, centred at right ascension 180 and declination 35. The longitude and latitude lines are spaced at 30° intervals.

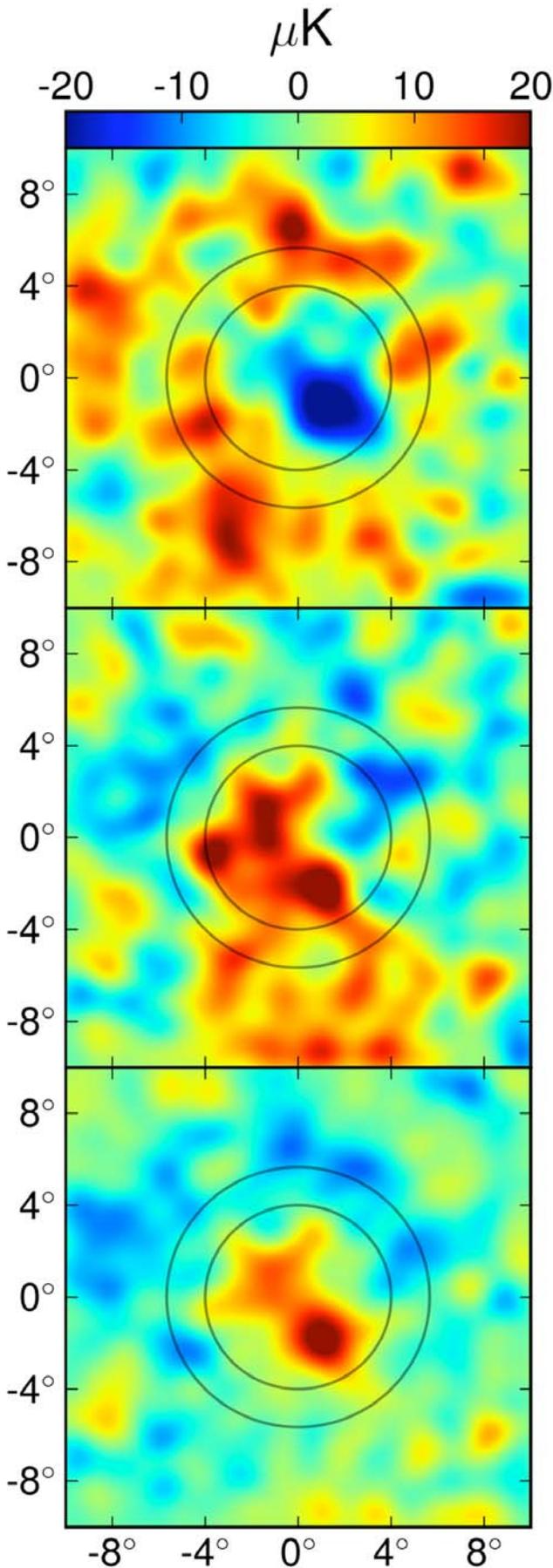

Figure 2: We stack regions on the CMB corresponding to supervoid and supercluster structures identified in the Sloan Digital Sky Survey. We averaged CMB cutouts around 50 supervoids (top) and 50 superclusters (middle), and differenced these two samples (bottom). The individual cutouts from the CMB were aligned vertically in the image based on the measured orientations of the clusters and voids, but we do not scale or apply weights to the images. Although our statistical analysis uses the raw image, for this figure we smooth the images with a Gaussian kernel with width 0.5°. A hot spot and a cold spot are immediately recognizable in the cluster and void stacks, respectively, with a characteristic radius of 4°, corresponding to spatial scales of 100 $h^{-1}$ Mpc. The inner circle (4° radius) and equal-area outer ring mark the extent of the compensated filter used in our analysis. The measured signal from these large structures is consistent with the ISW effect. There is a tantalizing hint of a hot ring around the cold spot. The observed morphology is consistent with the 'cosmic web'[30] picture in which voids are typically surrounded with 'walls' of higher density regions, while clusters fade gradually into the surrounding with filaments originating from them. Given the somewhat arbitrary rotations of each image in the stack, and the noise level, small-scale features should be interpreted cautiously.

# Supplementary Information

## Galaxy sample

Our galaxy sample was selected from SDSS[1] Data Release 6 with identical criteria to the MegaZ LRG catalogue[2], producing a catalogue of 1.2 million sources with redshifts 0.45<z<0.75. The photometric redshifts of the MegaZ catalogue are calibrated with spectroscopy of 3000 galaxies spanning the redshift range[3] and have characteristic 1σ errors of Δz=.05. This is a standard galaxy sample now used in many Large-Scale Structure studies[4,5], including ISW analyses[6]. For this work, we extended the LRG selection over a 20% greater sky area than the original MegaZ catalogue (based on DR4). To extrapolate the calibrated photometric redshifts to the new objects, we perform a 4-dimensional nearest-neighbour match in the g,r,i,z filter photometry space, using the MegaZ catalogue as the training set. In our algorithm, the nearest neighbours are found using a kd-tree data structure, and redshifts are assigned by a weighted mean of all galaxies within a 0.1 magnitude radius. The photo-z assignments are made over the entire galaxy sample, including those in the original MegaZ catalogue, thus providing a consistent photo-z estimate over the entire area. Effectively, this procedure smooths the redshift distribution, but we find that the resulting distribution is consistent with the original.

## Statistical details

Our detection is made from combining the mean signal from 50 supervoid regions and 50 supercluster regions on the CMB. The detection amplitudes measured in a 4° radius filter of these separate stacks are listed in table 1. The significance of our detection was modelled with two Monte Carlo runs, as discussed in the letter. These produce consistent confidence limits to 2%, and are consistent with a Gaussian distribution (see Figure 1). We tested the robustness of our detection by varying the analysis parameters. We varied the detection filter radius between 3° and 5° and measured the signal from the most significant 30 and 70 voids and clusters. (The main analysis was carried out with 50 each). The results are summarized in tables 2 and 3.

| | ΔT (μK) | ΔT/σ |
|---|---|---|
| Supervoids | -11.3 | 3.7 |
| Superclusters | 7.9 | 2.6 |

Table 1. The detection amplitudes in the stack of 50 supervoid regions and 50 supercluster regions measured in a 4° radius compensated filter.

| Radius (deg) | ΔT (μK) | ΔT/σ |
|---|---|---|
| 3.0° | 8.4 | 3.5 |
| 3.5 | 9.3 | 4.0 |
| 4.0 | 9.6 | 4.4 |
| 4.5 | 9.2 | 4.4 |
| 5.0 | 7.8 | 3.8 |

Table 2. The detection signifiance of the combined supervoid and supercluster stack measured with filters of various scales.

| Number in stack | ΔT (μK) | ΔT/σ |
|---|---|---|
| 30 | 11.1 | 4.0 |
| 70 | 5.4 | 2.8 |

Table 3. The combined signal measured from top 30 and 70 supervoid and supercluster samples.

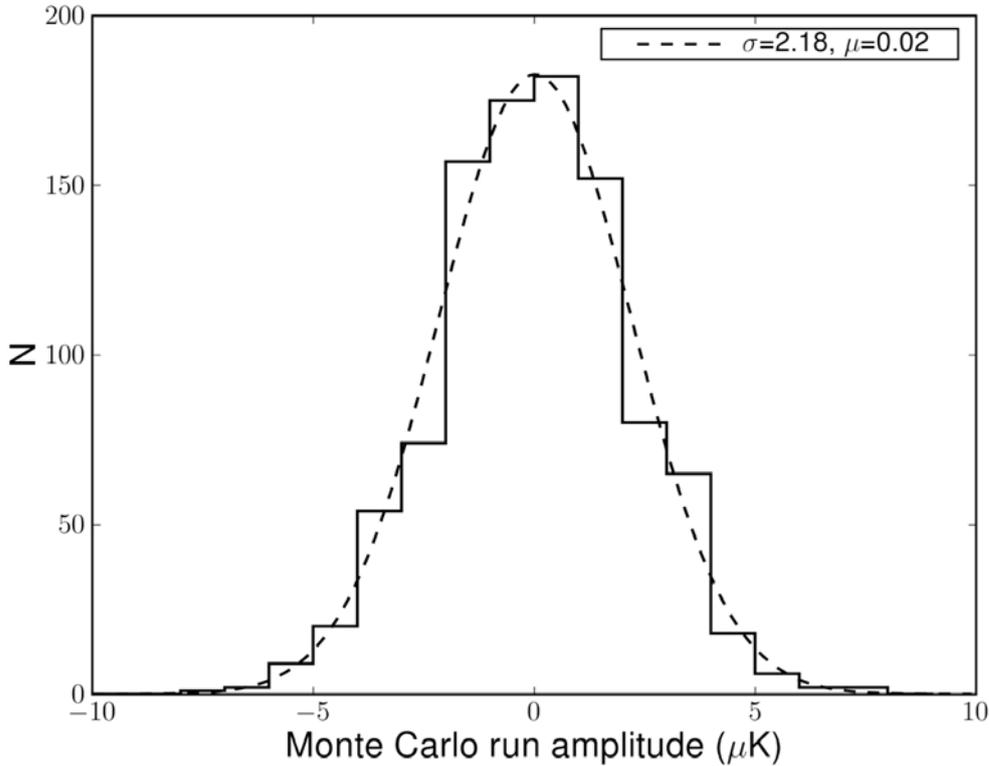

Figure 1. A histogram of amplitudes obtained by measuring the combined supervoid and supercluster signal on 1000 model CMB realizations. The best-fitting Gaussian is overplotted.

## ISW estimation using the Millennium Simulation

Here we discuss our estimation of the magnitude of the ISW (or Rees-Sciama) effect by sending photons[7] through the potential field of the Millennium simulation[8]. This is an *N*-body dark-matter cosmological simulation that has a box size of 500 $h^{-1}$ Mpc, has exquisite mass and spatial resolution, and was run using a concordance ΛCDM cosmology, including 75% dark energy. Ray tracing involves integrating the time-derivative of the potential, $\dot{\Phi}$, over light paths, i.e. summing up that derivative in columns in the *x*, *y*, and *z* directions. We downloaded publicly available data cubes of the density estimated using a clouds-in-cell algorithm, sampled on a $256^3$ grid, at redshifts *z*=0 and 0.02. We then measured the potential at each grid cell using a Fast Fourier Transform. To get $\dot{\Phi}$, we use two different methods. The first is a linear method, temporally extrapolating the potential at each grid cell at *z*=0 based on the linear growth factor. The second, 'full' method measures the time derivative using the potential in the previous, *z*=0.02 timestep.

Before discussing the ray-trace through the whole simulation, first we will discuss a small box 40 $h^{-1}$ Mpc on a side, the same volume that was analyzed in the Aspen-Amsterdam Void-Finder Comparison Project (AAVFCP)[9]. This volume contains one of the largest voids in the simulation; the void is a region 10-20 $h^{-1}$ Mpc in radius where the density stays below 0.2 times the mean. Figure 2 shows that the nonlinear contribution to the ISW signal through this relatively small volume can reach the same order as the linear signal, amplifying the cooling effect of the void.

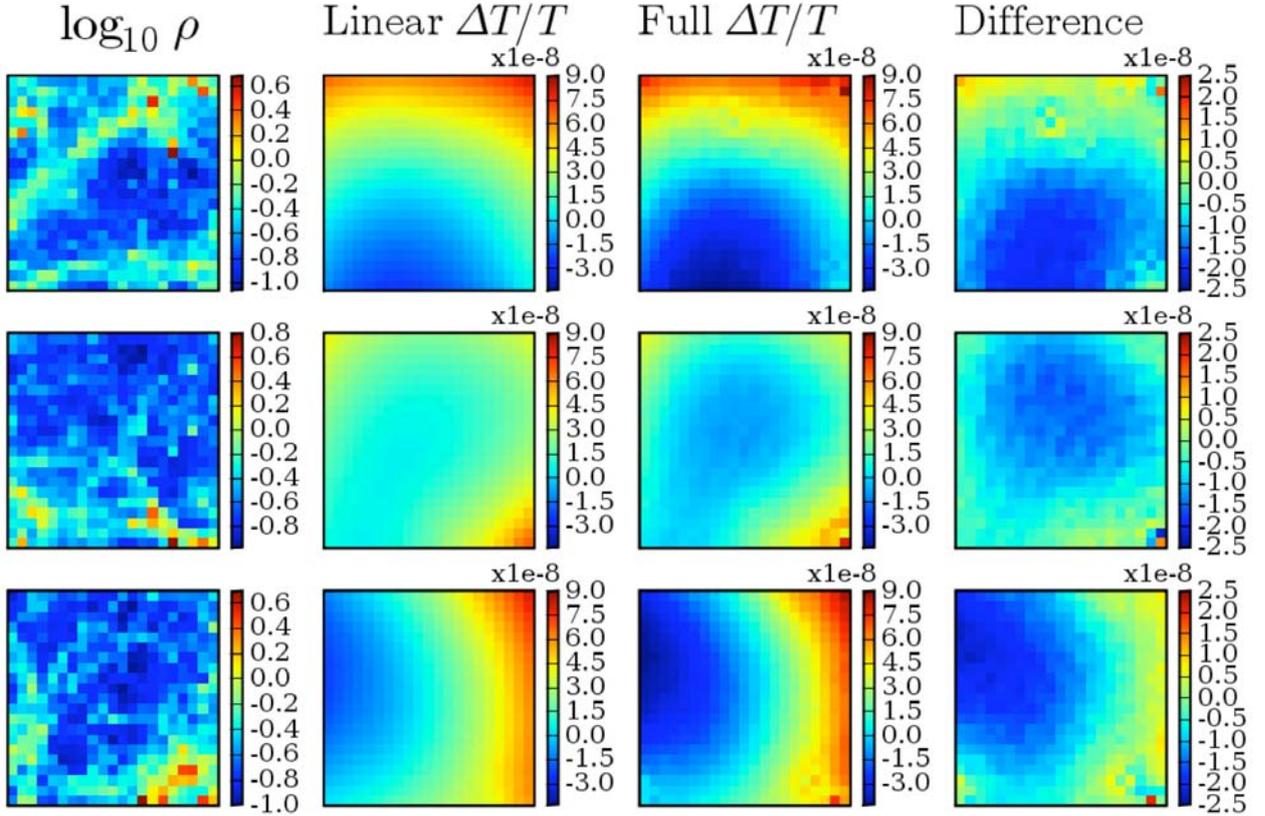

Figure 2 The ISW signal stacked in the three cardinal directions (one per row) through a cube 40 $h^{-1}$ Mpc on a side that contains a void (not supervoid), of radius 10-20 $h^{-1}$ Mpc. The void comes from the Millennium simulation. The first column shows the logarithm of the density averaged in columns through the cube. The second and third columns show the linear (extrapolated from a single snapshot of the simulation), and the full (using two adjacent snapshots) ISW signals. The right column shows their difference, the nonlinear contribution to the ISW signal; its contribution to the signal from this void can be significant, and amplifies it.

In Figure 3, we show the ISW signal through the whole simulation. Looking in each Cartesian direction, there are generally only one or two large cold spots, and one or two large hot spots, in the simulation. These presumably correspond to supervoids and superclusters. The 'large void' shown in Figure 2, which is projected as the white square in each dimension, has an entirely negligible signal in Figure 3, and thus does not qualify as a supervoid. About 36 Millennium volumes would fit in the LRG sample, so it is plausible that about that many give a measurable signal. In Figure 3, $\Delta T/T$ ranges from $-2.3 \times 10^{-6}$ to $2.2 \times 10^{-6}$, giving $\Delta T$ of -6.4μK to 6.1μK. Averaging in a 100 $h^{-1}$ Mpc-radius aperture, as we roughly do in the actual measurement, gives 4.2μK, within an order of magnitude of what we measure, 11μK, but is 1.9σ lower. One factor that tends to reduce our estimate unrealistically is the finite, rather small box size of the simulation (effectively, even smaller than 500 $h^{-1}$ Mpc, because of the periodic boundary conditions). Larger, and perhaps more numerous, supervoids and superclusters could form in the real universe. On the other hand, our measurement was at redshift 0, not at 0.5, when the ISW effect would be a bit weaker, since dark energy is not thought to have been as dominant then.

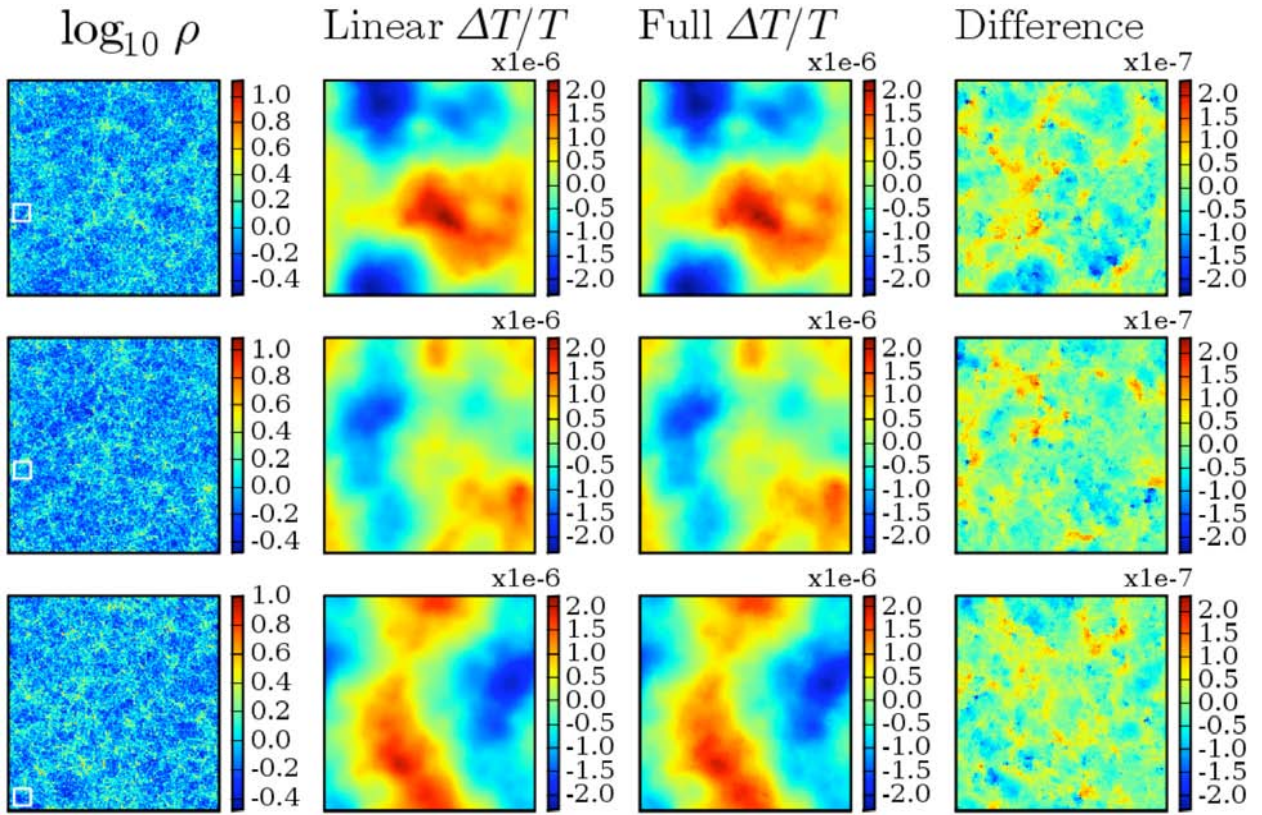

Figure 3. The ISW signal stacked in the three cardinal directions (one per row) through the whole 500 $h^{-1}$ Mpc volume of the Millennium simulation. The region in Figure 2 is projected into the white squares in the density (leftmost) column. There are generally only one or two large hot and cold spots per frame (the boundary conditions are periodic); these presumably correspond to cold spots. Here, the nonlinear ('difference') contribution is quite weak, but there is a hint that is skewed coldward, even in hot spots, as predicted analytically[10].

The nonlinear contribution to the signal in Figure 3 is weak, typically well under an order of magnitude smaller than the linear signal. This supports the idea that what we detect in the Letter is the linear ISW effect; booming superstructures wash out signals from non-linear structures such as that shown in Figure 2.

## Void and cluster finding

Here we discuss some details of how we found voids and clusters in the SDSS LRG sample. The term 'density depression' as used in the letter to describe a void could do with further definition. In 2D, if density were represented as height, the density depressions ZOBOV[11] finds would correspond to catchment basins[12]. Large voids can include multiple depressions, joined together to form a most-probable extent. This requires judging the significance of a depression; for this, we use its density contrast, comparing against density contrasts of voids from a uniform Poisson point sample. Most of the voids in our catalogue consist of a single depression. As stated in the letter, we used the same algorithm run on the inverse of the density, VOBOZ[13], to detect clusters; even fewer clusters consist of multiple density enhancements.

We estimated the density of the galaxy sample in three dimensions, converting redshift to distance according to WMAP5[14] cosmological parameters. To correct for the variable selection function, we normalized the galaxy densities to have the same mean in 100 equally spaced distance bins. This also removed dependence on the redshift-distance mapping.

We handled survey boundaries by putting fake buffer galaxies around the sample in each dimension, and then discarding any real galaxies with Voronoi neighbours in the buffer. The buffer around the sky footprint was 1° thick, and randomly sampled at thrice the mean density. We handled holes by filling then with random fake galaxies at the mean density. The hole galaxies comprised about 1/300 of the galaxies used to find voids and clusters. The redshifts of hole and buffer galaxies were randomly sampled from the real galaxies. The near and far buffers in the redshift dimension were meshes with maximum spacing of 1°, put slightly (at a distance corresponding to 1°) inside the redshift range. All of this ensured that no galaxies were analyzed for void or cluster membership whose density and set of neighbours could be affected by undetected galaxies outside the sample.

Table 4: Catalogue of the 50 highest-significance ZOBOV (super)voids used for our measurement. $z$ is the redshift of the lowest-density galaxy in the void. Voids were excluded that had over 10% overlap with a hole (this excluded only one void from this list), and that were within 2.5° of the survey edge. RA and Dec are the right ascension and declination of the void centers-of-volume, i.e. the average positions of the galaxies on the sky, weighted by the volume of their Voronoi cell (after correcting for the selection-function variation with redshift). Rsky and Rmax are the mean and maximum distances on the sky, in degrees, between galaxies in a void and their center of volume. Vol is the sum of the volumes of Voronoi cells (with no selection-function correction) around galaxies in the void, in $h^{-3}$ Mpc$^3$. d_all, d_neg, and d_min are three different measures of void overdensity $\delta=\rho-1$, where $\rho$ is the density (with the selection-function correction), in units of the mean. d_all uses all galaxies in each void; for a few voids it is positive because overdense galaxies can be included at the edge. d_neg is the average of all underdense galaxies in the void. d_min is the minimum density of galaxies in the void. Contr is the density contrast, the ratio of the density at which the void would percolate into another void to d_min. Prob is the probability that a void has that density contrast in a Poisson sampling of points, according to Eqn. (1) of ref. 11.

```
z     RA     Dec   Rsky Rmax Vol     d_all   d_neg  d_min  Contr Prob
0.451 197.82 55.62 1.91 3.95 1.3e+06 -0.003  -0.376 -0.827 3.025 9.8e-08
0.449 145.34 57.67 2.58 4.30 1.7e+06 -0.163  -0.424 -0.856 2.890 5.4e-07
0.452 237.32 48.09 3.11 7.16 3.2e+06 -0.199  -0.413 -0.869 2.869 6.9e-07
0.464 205.28 25.16 1.65 3.29 1.1e+06 -0.033  -0.343 -0.818 2.813 1.4e-06
0.449 214.31 36.69 2.15 4.06 1.2e+06 -0.016  -0.389 -0.826 2.764 2.4e-06
0.665 161.85  8.79 2.64 5.48 7.7e+06 -0.193  -0.397 -0.837 2.735 3.3e-06
0.684 203.30 62.99 2.85 6.39 9.5e+06 -0.123  -0.371 -0.817 2.566 2.0e-05
0.547 234.55  5.89 1.25 2.41 7.3e+05 -0.143  -0.363 -0.820 2.414 8.7e-05
0.537 200.91 49.97 1.49 2.52 1.2e+06 -0.084  -0.372 -0.815 2.394 1.0e-04
0.437 194.75 12.14 2.72 6.70 3.5e+06 -0.090  -0.384 -0.807 2.367 1.3e-04
0.603 189.23 22.09 1.79 4.31 2.8e+06 -0.165  -0.362 -0.802 2.363 1.4e-04
0.448 208.74 50.97 2.19 4.71 1.5e+06 -0.223  -0.468 -0.869 2.337 1.8e-04
0.648 220.28 20.05 3.19 7.21 1.1e+07 -0.197  -0.407 -0.829 2.336 1.8e-04
0.446 187.83  0.16 2.47 5.53 2.1e+06 -0.093  -0.383 -0.786 2.332 1.8e-04
0.451 153.80 46.69 2.46 5.68 1.7e+06  0.005  -0.367 -0.791 2.314 2.1e-04
0.581 150.69  1.62 1.57 3.47 1.1e+06 -0.088  -0.351 -0.811 2.300 2.4e-04
0.667 224.79 13.70 2.64 5.60 5.7e+06 -0.194  -0.431 -0.858 2.295 2.5e-04
0.471 166.21 20.27 1.68 3.77 1.5e+06 -0.156  -0.426 -0.834 2.281 2.9e-04
0.447 200.59  8.24 1.84 3.98 1.3e+06 -0.018  -0.357 -0.784 2.275 3.0e-04
0.672 193.91 25.02 3.26 6.34 1.0e+07 -0.316  -0.441 -0.837 2.273 3.1e-04
0.446 226.00  3.06 2.30 4.87 2.1e+06 -0.211  -0.414 -0.838 2.263 3.3e-04
0.523 161.29 40.97 1.81 4.16 2.0e+06 -0.139  -0.386 -0.828 2.233 4.3e-04
0.449 120.03 33.87 2.07 3.75 1.1e+06 -0.123  -0.375 -0.809 2.206 5.4e-04
0.522 152.95 35.51 0.71 1.24 1.1e+05  0.396  -0.255 -0.647 2.204 5.5e-04
0.495 206.37 25.09 1.48 2.83 1.1e+06 -0.141  -0.362 -0.804 2.186 6.3e-04
0.623 149.48  9.41 2.01 3.82 5.2e+06 -0.059  -0.333 -0.758 2.184 6.5e-04
0.662 133.16 11.75 2.56 4.96 6.5e+06 -0.131  -0.365 -0.779 2.175 6.9e-04
0.561 220.74 40.14 1.38 2.69 9.0e+05 -0.085  -0.336 -0.771 2.167 7.4e-04
0.445 209.31 29.22 1.79 3.55 6.3e+05  0.020  -0.307 -0.711 2.166 7.5e-04
0.566 161.64 40.88 1.76 3.87 1.6e+06 -0.089  -0.379 -0.784 2.151 8.5e-04
0.442 222.07 27.70 2.07 5.37 2.1e+06 -0.139  -0.402 -0.794 2.147 8.7e-04
0.450 123.50 42.83 2.15 6.23 1.6e+06 -0.222  -0.419 -0.821 2.135 9.6e-04
0.584 185.94 35.48 1.59 3.84 2.0e+06 -0.136  -0.381 -0.795 2.102 1.2e-03
0.595 155.50 14.52 1.24 2.80 8.6e+05 -0.023  -0.299 -0.726 2.097 1.3e-03
0.555 219.59 21.44 1.17 2.50 5.6e+05 -0.118  -0.360 -0.772 2.092 1.3e-03
0.511 218.98 24.31 1.67 4.51 1.6e+06 -0.099  -0.367 -0.771 2.076 1.5e-03
0.558 194.76 25.09 1.01 1.94 5.2e+05 -0.190  -0.331 -0.764 2.065 1.7e-03
0.518 188.05 22.50 1.51 3.98 6.7e+05 -0.161  -0.376 -0.783 2.065 1.7e-03
0.476 148.15 24.85 1.75 3.96 1.9e+06 -0.032  -0.344 -0.759 2.063 1.7e-03
0.449 131.34 44.12 2.38 5.00 2.1e+06 -0.074  -0.370 -0.798 2.062 1.7e-03
0.466 140.01 20.20 2.15 4.84 2.3e+06 -0.144  -0.416 -0.823 2.058 1.7e-03
0.504 218.83 17.75 1.86 3.60 1.2e+06 -0.098  -0.360 -0.789 2.048 1.9e-03
0.555 226.55 19.54 1.09 2.45 4.0e+05  0.084  -0.323 -0.726 2.048 1.9e-03
0.675 181.09 17.77 2.61 5.02 6.0e+06 -0.333  -0.474 -0.820 2.043 2.0e-03
0.466 244.87 11.30 1.67 4.90 1.4e+06  0.104  -0.318 -0.718 2.035 2.1e-03
0.635 132.33 55.55 2.15 4.73 5.9e+06 -0.034  -0.333 -0.746 2.028 2.2e-03
0.525 181.30  8.54 1.83 3.82 1.2e+06 -0.108  -0.365 -0.810 2.014 2.4e-03
0.465 187.79  9.18 1.95 4.40 1.8e+06 -0.130  -0.376 -0.783 2.013 2.4e-03
0.514 194.07 60.29 1.32 2.74 7.0e+05 -0.126  -0.356 -0.786 2.010 2.5e-03
0.498 247.89 36.48 1.43 3.37 1.2e+06  0.057  -0.332 -0.716 2.007 2.6e-03
```

Table 5: Catalogue of the 50 highest-significance VOBOZ (super)clusters used for our measurement. For the RA and Dec here, we performed a simple average of the galaxy sky positions, not weighting by volume. The other fields are the same as, or analogous to, the fields in the void catalogue. The probabilities come from Eqn. (1) of ref. 13. Clusters were excluded that had over 10% overlap with a hole (a criterion that did not affect this list), and that were within 2.5° of the survey edge. Additionally, we excluded one cluster because its position on the sky was within 0.5° of an included void that was both bigger and more significant (judging by its 'Prob'), and had larger volume.

```
z      RA     Dec    Rsky Rmax Vol      d_all  d_pos d_max  Contr Prob
0.467 123.50 39.43  0.58 1.53 7.4e+04   0.435  0.663 17.244 9.987 5.3e-04
0.479 139.85 51.10  0.48 1.08 5.8e+04   0.517  1.095 32.581 9.879 5.5e-04
0.513 163.38 56.83  0.33 0.77 2.5e+04   0.335  0.653 15.004 9.846 5.6e-04
0.527 190.66 32.50  0.58 1.26 1.7e+05  -0.148  0.444 12.537 9.802 5.7e-04
0.477 193.61  5.26  0.43 1.05 7.4e+04  -0.007  1.217 16.160 9.788 5.8e-04
0.567 181.66 37.53  0.25 0.49 3.2e+04   0.079  0.507 12.639 9.761 5.8e-04
0.457 128.43 26.91  0.55 1.47 9.5e+04   0.241  0.902 21.402 9.685 6.0e-04
0.470 181.60 30.61  0.43 1.15 2.8e+04   0.697  1.031 27.420 9.647 6.1e-04
0.545 231.26 41.47  0.54 1.38 1.5e+05  -0.130  0.554 12.929 9.623 6.2e-04
0.469 229.89 49.76  0.42 0.81 5.5e+04   0.086  0.732 14.896 9.465 6.7e-04
0.516 223.93 13.24  0.50 1.33 6.8e+04   0.211  0.878 18.101 9.430 6.8e-04
0.627 212.83  6.01  0.69 1.87 3.1e+05   0.017  0.666 12.175 9.427 6.8e-04
0.502 216.69 25.79  0.53 1.28 1.0e+05  -0.052  0.794 13.889 9.426 6.8e-04
0.470 170.14 62.45  0.69 1.68 1.2e+05  -0.178  0.524 11.489 9.402 6.9e-04
0.453 182.47 24.60  0.50 0.95 3.7e+04   0.151  0.655 21.462 9.394 6.9e-04
0.457 204.61 35.75  0.49 1.07 8.8e+04  -0.106  0.634 14.368 9.384 6.9e-04
0.556 241.02 21.64  0.55 1.28 9.0e+04   0.110  0.822 16.398 9.376 7.0e-04
0.535 201.73 53.32  0.52 1.30 1.2e+05   0.568  0.905 18.928 9.336 7.1e-04
0.458 144.20 45.48  0.73 2.25 2.7e+05   0.026  0.714 18.729 9.261 7.3e-04
0.489 203.87 29.60  0.23 0.53 2.6e+04   0.241  0.662 17.552 9.214 7.5e-04
0.465 220.25 33.81  0.51 1.32 1.3e+05   0.055  0.607 16.502 9.196 7.6e-04
0.463 121.88 45.65  0.42 1.21 8.1e+04   0.014  0.864 10.562 9.180 7.6e-04
0.522 237.75 14.16  0.39 0.90 3.0e+04   0.167  0.868 16.437 9.178 7.6e-04
0.490 175.99 18.66  0.36 0.75 4.5e+04   0.380  0.545 20.100 9.112 7.9e-04
0.458 186.72 46.77  0.46 1.23 4.3e+04   0.159  0.725 11.661 9.073 8.0e-04
0.492 229.10 56.95  0.29 0.90 3.3e+04  -0.002  0.684 13.377 9.055 8.1e-04
0.476 185.00  9.07  0.44 0.89 6.2e+04   0.144  0.564 11.530 8.978 8.4e-04
0.570 241.00 15.90  0.49 0.90 1.1e+05   0.093  0.554 12.048 8.970 8.4e-04
0.547 230.81  8.42  0.38 0.89 5.9e+04   0.449  0.701 17.076 8.959 8.5e-04
0.552 220.95  1.55  0.47 1.01 6.8e+04   0.197  0.684 17.477 8.894 8.7e-04
0.464 202.75 52.54  0.38 0.76 3.2e+04   0.236  0.599 14.000 8.864 8.9e-04
0.541 139.05 21.13  0.53 1.76 1.8e+05  -0.200  0.663  8.276 8.810 9.1e-04
0.511 192.45  6.37  0.48 1.08 9.7e+04  -0.083  0.568  9.773 8.805 9.1e-04
0.581 131.91 17.07  0.43 1.12 1.1e+05  -0.016  0.804 14.227 8.801 9.1e-04
0.477 161.46 21.79  0.29 0.77 2.3e+04   0.003  0.682 14.179 8.772 9.3e-04
0.456 139.15 60.18  0.62 1.26 1.7e+05  -0.165  0.795  7.808 8.728 9.5e-04
0.463 175.20 12.20  0.47 1.34 8.4e+04   0.023  0.982 14.521 8.570 1.0e-03
0.470 124.25 20.87  0.41 0.94 4.5e+04   0.002  0.683 12.190 8.562 1.0e-03
0.529 161.07 20.11  0.40 1.02 6.2e+04   0.355  0.755 15.552 8.546 1.0e-03
0.487 166.07 11.86  0.35 0.63 5.2e+04  -0.177  0.807  8.596 8.501 1.1e-03
0.460 149.93 52.28  0.39 0.99 4.4e+04   0.400  0.672 15.365 8.488 1.1e-03
0.488 130.91 28.71  0.39 0.77 4.7e+04   0.289  0.763 17.963 8.424 1.1e-03
0.477 209.03 31.96  0.54 1.09 4.9e+04   0.379  0.789 23.096 8.422 1.1e-03
0.621 226.41  5.46  0.78 1.75 5.1e+05  -0.038  0.439 10.289 8.405 1.1e-03
0.448 145.39  0.63  0.81 1.85 3.6e+05   0.037  0.695 14.558 8.404 1.1e-03
0.541 199.69  0.03  0.44 1.20 9.5e+04   0.123  0.993 17.765 8.404 1.1e-03
0.503 242.77 25.75  0.33 0.64 3.7e+04   0.426  0.994 17.007 8.401 1.1e-03
0.566 142.76 33.35  0.59 1.46 1.4e+05  -0.277  0.491  9.314 8.363 1.1e-03
0.549 148.58 11.57  0.50 1.19 1.1e+05   0.003  0.612  9.284 8.357 1.1e-03
0.535 158.60 44.77  0.55 1.22 1.1e+05   0.148  0.532 12.223 8.350 1.1e-03
```